\documentclass[slac_one]{revtex4}
\usepackage{graphicx}
\usepackage{fancyhdr}
\usepackage{epsfig,amsmath,amssymb}

\newcommand{\beq}{\begin{equation}}
\newcommand{\eeq}{\end{equation}}
\newcommand{\bea}{\begin{eqnarray}}
\newcommand{\eea}{\end{eqnarray}}
\newcommand{\Fig}[1]{Fig.\,\ref{#1}}

\newcommand{\Eq}[1]{Eq.\,(\ref{#1})}

\newcommand{\f}{\frac}

\newcommand{\as}{\alpha_s}

\newcommand{\MW}{M_{\scriptscriptstyle W}}
\newcommand{\MZ}{M_{\scriptscriptstyle Z}}
\newcommand{\mc}{m_c}
\newcommand{\mb}{m_b}
\newcommand{\mt}{m_t}

\newcommand{\muh}{\mu_{\scriptscriptstyle W}}

\newcommand{\muc}{\mu_c}
\newcommand{\mub}{\mu_b}

\newcommand{\GF}{G_F}
\newcommand{\BR}{{\cal{B}}}
\newcommand{\GeV}{{\rm \ GeV}}
\newcommand{\TeV}{{\rm \ TeV}}
\newcommand{\MSbar}{\overline{\rm MS}}
\newcommand{\re}{{\rm Re}}
\newcommand{\im}{{\rm Im}}
\newcommand{\ord}{{\cal O}}

\newcommand{\sL}{{\scalebox{0.6}{$L$}}}

\newcommand{\Ktopinunu}{K^+ \to \pi^+ \nu \bar{\nu}}
\newcommand{\stodnunu}{s \to d \nu \bar{\nu}}

\newcommand{\KLtopinunu}{K_L \to \pi^0 \nu \bar{\nu}}

\newcommand{\Ktopienu}{K^+ \to \pi^0 e^+ \nu}
\newcommand{\Ktopinunus}{K \to \pi \nu \bar{\nu}}

\newcommand{\Pc}{P_c}
\newcommand{\dPuc}{\delta P_{c}}

\begin{document}

\title{{\small{2005 ALCPG \& ILC Workshops - Snowmass,
U.S.A. \hfill FERMILAB-PUB-05-526-T; ZU-TH 24/05; hep-ph/0512007}}\\ 
\vspace{12pt} 
Rare {\boldmath $\Ktopinunus$} Decays} 

%

\author{Ulrich Haisch}
\affiliation{Theoretical Physics Department, Fermilab, Batavia, IL
60510, U.S.A.} 
\affiliation{Institut f\"ur Theoretische Physik, Universit\"at
Z\"urich, CH-8057 Z\"urich, Switzerland} 

\begin{abstract}
We present a concise review of the theoretical status of rare
$\Ktopinunus$ decays in the standard model (SM). Particular attention
is thereby devoted to the recent calculation of the
next-to-next-to-leading order (NNLO) corrections to the charm quark 
contribution of $\Ktopinunu$, which removes the last relevant
theoretical uncertainty from the $\Ktopinunus$ system.     
\end{abstract}

\maketitle

\thispagestyle{fancy}


\section{\label{sec:introduction}INTRODUCTION}

The rare processes $\Ktopinunu$ and $\KLtopinunu$ play an outstanding
role in the field of flavor changing neutral current transitions. The
main reason for this is their unmatched theoretical cleanliness and
their large sensitivity to short-distance (SD) effects arising in the
SM and its innumerable extensions. As they offer a very precise
determination of the unitarity triangle (UT) \cite{UT}, a comparison
of the information obtained from the $\Ktopinunus$ system with the one
from $B$-decays provides a completely independent and therefore
critical test of the Cabibbo-Kobayashi-Maskawa (CKM) mechanism. Even
if these $K$- and $B$-physics predictions agree, the $\Ktopinunus$
transitions will play a leading, if not the leading part in
discriminate between different extensions of the SM
\cite{Buras:2004uu}, as they allow to probe effective scales of new 
physics operators of up to a several $\!\! \TeV$ or even higher in a
pristine manner.   

\section{\label{sec:basic}BASIC PROPERTIES OF {\boldmath $\Ktopinunus$}}

The striking theoretical cleanliness of the $\Ktopinunus$ decays is
linked to the fact that, within the SM, these processes are mediated
by electroweak (EW) amplitudes of $\ord (\GF^2)$, which exhibit a hard
Glashow-Iliopoulos-Maiani cancellation of the form    
\beq \label{eq:Aq}
{\cal A}_q (\stodnunu) \propto \lambda_q m_q^2 \propto \begin{cases}
\mt^2 ( \lambda^5 + i \lambda^5 ) \, , & \hspace{1mm} q = t \, , \\
\mc^2 ( \lambda + i \lambda^5 ) \, , & \hspace{1mm} q = c \, , \\
\Lambda_{\rm QCD}^2 \lambda \, , & \hspace{1mm} q = u \, , 
\end{cases}   
\eeq
where $\lambda_q = V^\ast_{qs} V_{qd}$ denotes the relevant CKM factors
and $\lambda = | V_{us} | = 0.2248$ is the Cabibbo angle. This peculiar
property implies that the corresponding rates are SD dominated, while
long-distance (LD) effects are highly suppressed. A related important
feature, following from the EW structure of the SM amplitudes as well,
is that the $\Ktopinunus$ modes are governed by a single effective
operator, namely 
\beq \label{eq:Qv}
Q_\nu = \left (\bar s_\sL \gamma_\mu d_\sL \right ) \left ( \bar
{\nu}_\sL \gamma^\mu {\nu}_\sL \right ) \, ,  
\eeq
which consists of left-handed fermion fields only. The required
hadronic matrix elements of $Q_\nu$ can be extracted, including
isospin breaking corrections \cite{Marciano:1996wy}, directly from the
well measured leading semileptonic decay $\Ktopienu$. 

After summation over the three lepton families the SM branching ratios
for $\Ktopinunus$ can be written as \cite{Buchalla:1993wq,
Buchalla:1998ba, Isidori:2005xm, Buras:2005gr}  
\beq \label{eq:BREQ}
\begin{split} 
\BR (\Ktopinunu) = ( 5.04 \pm 0.17 ) \left [ \left ( \f{\im
\lambda_t}{\lambda^5} X \right )^2 + \left ( \f{\re 
\lambda_t}{\lambda^5} X + \f{\re \lambda_c}{\lambda} \left (\Pc +
\dPuc \right ) \right )^2 \right ] \times 10^{-11} \, , \\ 
\BR (\KLtopinunu) = ( 2.20 \pm 0.07 ) \, \left ( \f{\im
\lambda_t}{\lambda^5} X \right )^2 \times 10^{-10} \, . \hspace{30mm} 
\end{split}
\eeq
The top quark contribution $X = 1.464 \pm 0.041$ \cite{Buras:2005gr}
accounts for $63 \%$ and almost $100 \%$ of the total rates. It is
known through next-to-leading order (NLO) \cite{Buchalla:1998ba, X},
with a scale uncertainty of slightly less than $1 \%$. In
$\Ktopinunu$, corrections due to internal charm quarks and subleading
effects, characterized by $\Pc$ and $\dPuc$, amount to moderate $33
\%$ and a mere $4 \%$. Both contributions are negligible in the case
of the $\KLtopinunu$ decay, which by virtue of \Eq{eq:Aq} is purely CP
violating in the SM.         

\section{\label{sec:developments} RECENT DEVELOPMENTS IN {\boldmath
$\Ktopinunu$}} 

\begin{figure}[!t]
\begin{center}
\scalebox{0.45}{\includegraphics{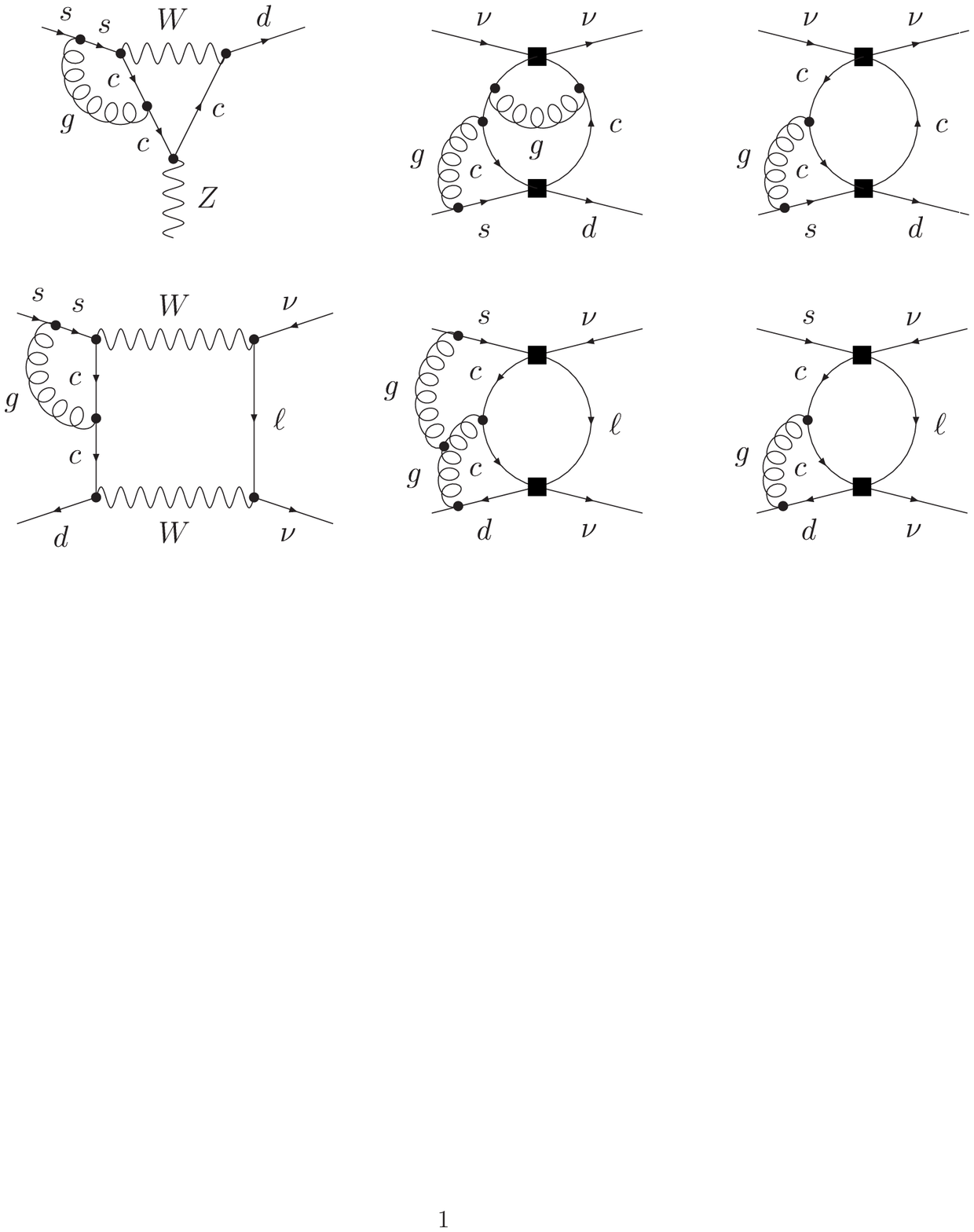}}
\rotatebox{270}{\scalebox{0.30}{\includegraphics{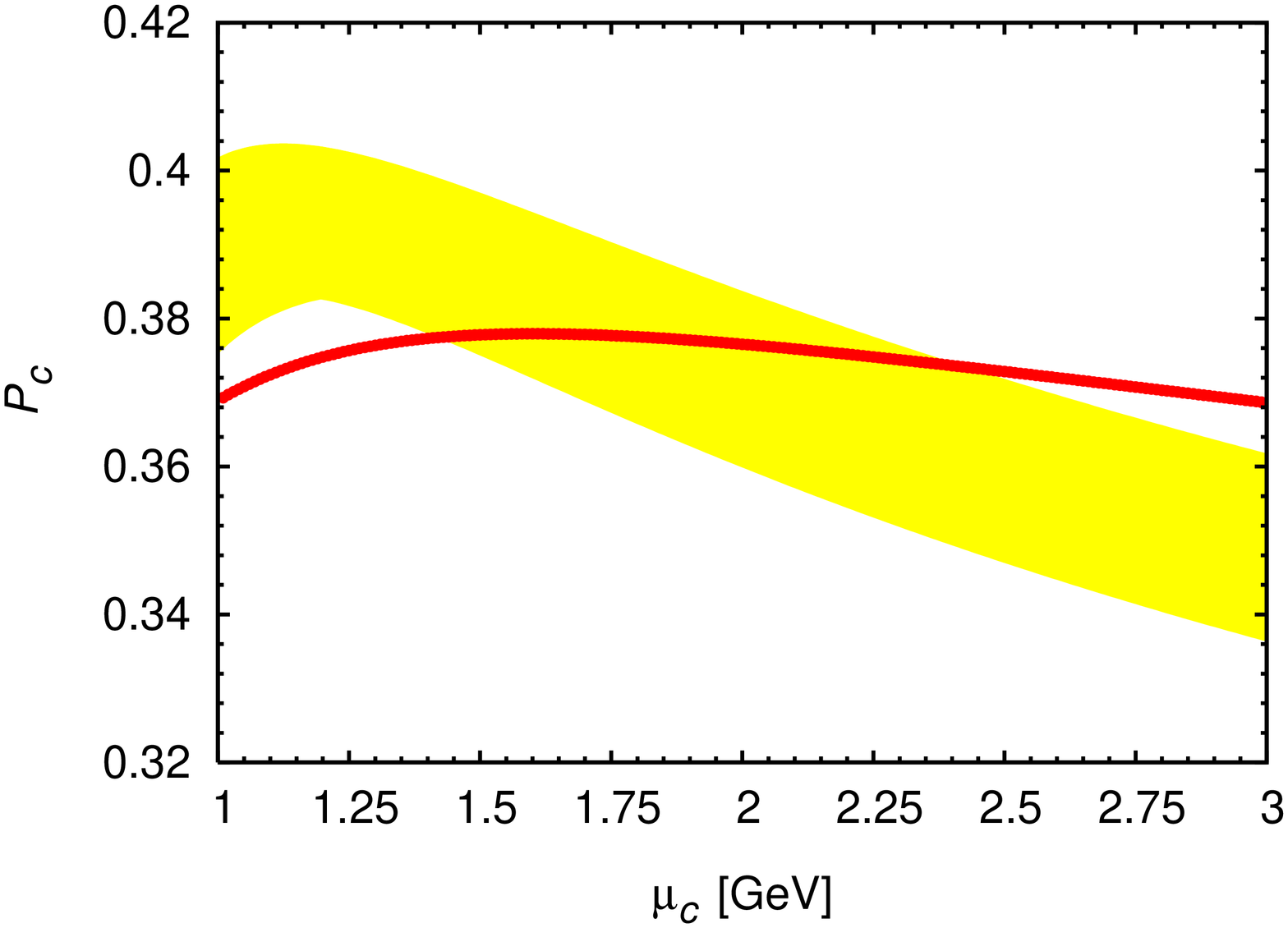}}}
\end{center}
\caption{Left panel: Examples of diagrams appearing in the full SM
(left column), describing the mixing of operators (center column) 
and the matrix elements (right column) in the $Z$-penguin (upper row)
and the electroweak box (lower row) sector. Right panel: $\Pc$ as a
function of $\muc$ at NLO (yellow band) and NNLO (red band). The width
of the bands reflects the theoretical uncertainty due to higher order
terms in $\as$ that arise in the calculation of $\as (\muc)$ from $\as
(\MZ)$.}         
\label{fig:fig1}
\end{figure}

Two subleading effects, namely the SD contributions of dimension-eight
charm quark operators and genuine LD corrections due to up quark loops
have been calculated recently \cite{Isidori:2005xm}. Both
contributions can be effectively included by $\dPuc = 0.04 \pm 0.02$ 
in \Eq{eq:BREQ}. Numerically they lead to an enhancement of $\BR    
(\Ktopinunu)$ by about $7 \%$. The quoted residual error of $\dPuc$
can in principle be reduced by means of dedicated lattice QCD
computations \cite{Isidori:2005tv}.    

The main components of the state-of-the-art calculation of $\Pc$
\cite{Buras:2005gr}, are $i)$ the $\ord (\as^2)$ matching corrections
to the relevant Wilson coefficients arising at $\muh = \ord (\MW)$,
$ii)$ the $\ord (\as^3)$ anomalous dimensions describing the mixing of
the dimension-six and -eight operators, $iii)$ the $\ord (\as^2)$
threshold corrections to the Wilson coefficients originating at $\mub
= \ord (\mb)$, and $iv)$ the $\ord (\as^2)$ matrix elements of some of
the operators emerging at $\muc = \ord (\mc)$. To determine the
contributions of type $i)$, $iii)$ and $iv)$ one must calculate
finite parts of two-loop Green's functions in the full SM and in
effective theories with five or four flavors. Sample diagrams for
steps $i)$ and $iv)$ are shown in the left and right column of the
left panel in \Fig{fig:fig1}. Contributions of type $ii)$ are found by
calculating the divergent pieces of three-loop Green's functions with
operator insertions. Two examples of Feynman graphs with a double
insertion of dimension-six operators are displayed in the center
column of the left panel in \Fig{fig:fig1}.   

Conceptual new features of this NNLO computation are $a)$ the
non-vanishing contribution from the vector component of the effective
neutral-current coupling describing the interaction of neutrinos and  
quarks mediated by $Z$-boson exchange, $b)$ the appearance of closed
quark loops in gluon propagators, resulting in a novel dependence of
$\Pc$ on the top quark mass and in non-trivial matching corrections at
the bottom quark threshold, and $c)$ the presence of anomalous triangle
diagrams involving a top quark loop, two gluons and a $Z$-boson making
it necessary to introduce a Chern-Simons operator in order to obtain
the correct anomalous Ward identity of the axial-vector current. The
inclusion of such a term is also required to cancel the anomalous
contributions from triangle diagrams with a bottom quark loop. Since
all these effects were absent in the NLO renormalization group
analysis of $\Pc$ \cite{Buchalla:1993wq, Buchalla:1998ba}, their 
actual size cannot be estimated from the magnitude of the residual
scale uncertainties, but has to be determined by an explicit
calculation.  

The inclusion of the NNLO corrections removes essentially the entire
sensitivity of $\Pc$ on the unphysical scale $\muc$ and on higher
order terms in $\as$ that affect the evaluation of $\as (\muc)$ from
$\as (\MZ)$. This is explicated by the plot in the right panel of
\Fig{fig:fig1} and by the theoretical errors of the latest SM
predictions \cite{Buras:2005gr}        
\beq \label{eq:Pc}
\Pc = \begin{cases} 0.367 \pm 0.037_{\rm theory} \pm 0.033_{\mc} \pm
0.009_{\as} \, , & \hspace{1mm} \text{NLO} \, , \\ 0.371 \pm
0.009_{\rm theory} \pm 0.031_{\mc} \pm 0.009_{\as} \, , & \hspace{1mm}
\text{NNLO} \, .  
\end{cases} 
\eeq 
In obtaining these values the charm quark $\MSbar$ mass $\mc (\mc) =
(1.30 \pm 0.05) \! \GeV$ has been used. The residual error of 
$\Pc$ is now fully dominated by the parametric uncertainty from $\mc
(\mc)$. A better determination of $\mc (\mc)$ is thus an important
theoretical goal in connection with $\Ktopinunu$.         

Taking into account all the indirect constraints from the global UT 
fit \cite{Charles:2004jd}, the updated SM predictions of the two
$\Ktopinunus$ rates read
\beq \label{eq:BRSM}
\BR (\Ktopinunu) = \left ( 8.0 \pm 1.1 \right ) \times 10^{-11} \, ,
\hspace{10mm} \BR (\KLtopinunu) = \left ( 2.9 \pm 0.4 \right ) \times
10^{-11} \, .  
\eeq 
Owing to our still limited knowledge of $\lambda_t$, the reduction of
the theoretical error in $\Pc$ is at present not adequately reflected in
the error of $\BR (\Ktopinunu)$. However, given the expected
improvement in the extraction of the CKM elements and the foreseen
theoretical progress in the determination of $\mc (\mc)$, the
allowed ranges of the SM predictions for both $\BR (\Ktopinunu)$ and
$\BR (\KLtopinunu)$ should reach the $5 \%$ level, or better, by the
end of the decade.     

Experimentally the $\Ktopinunus$ modes are in essence unexplored up to
now. The AGS E787 and E949 Collaborations at Brookhaven observed the
decay $\Ktopinunu$ finding three events \cite{KpEX}, while there is
only an upper limit on $\KLtopinunu$, improved recently by the E391a
experiment at KEK-PS \cite{KLEX}. The corresponding numbers read  
\beq \label{eq:BREX}
\BR (\Ktopinunu) = \left ( 14.7^{+13.0}_{-8.9} \right ) \times
10^{-11} \, , \hspace{10mm} \BR (\KLtopinunu) < 2.86 \times 10^{-7}
\hspace{2mm} (90 \% \, {\rm  CL}) \, .
\eeq 
Within theoretical, parametric and experimental uncertainties,
the observed value of $\BR (\Ktopinunu)$ is fully consistent with the
present SM prediction given in \Eq{eq:BRSM}.   

\begin{figure}[!t]
\begin{center}
\scalebox{0.40}{\includegraphics{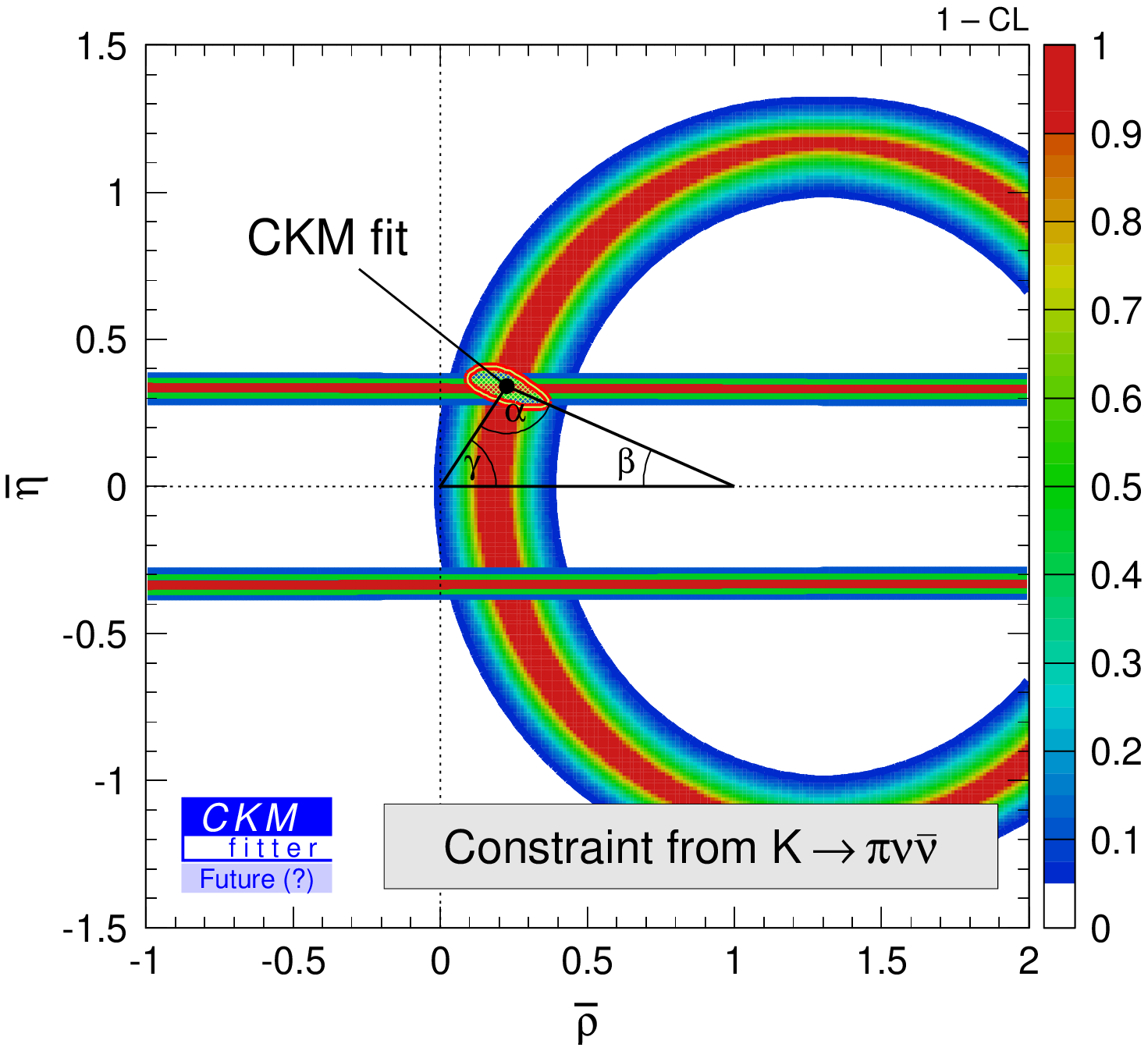}}
\scalebox{0.40}{\includegraphics{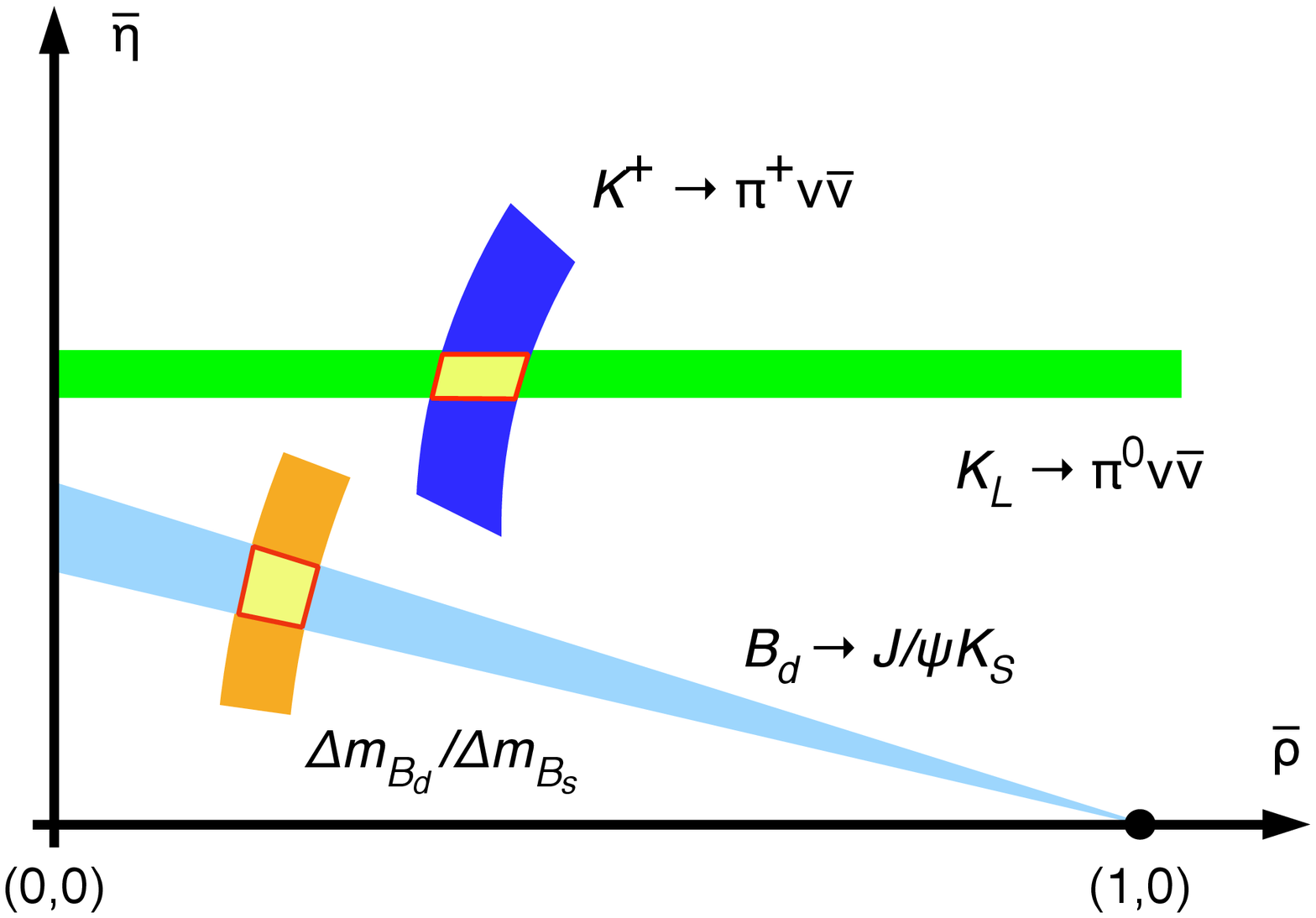}}
\end{center}
\caption{Left panel: UT from future measurements of $\BR (\Ktopinunu)$
and $\BR (\KLtopinunu)$ with an accuracy of $10 \%$. The constraint
from the present global CKM fit is overlaid. Right panel: Schematic
determination of the apex of the UT from either
$B_{d,s}$--$\bar{B}_{d,s}$ oscillations and the mixing induced CP
asymmetry in $B_d \to J/\psi K_s$ or from the $\Ktopinunus$ system in
the presence of non-minimal flavor violating new physics
contributions.}               
\label{fig:fig2}
\end{figure}

The impact of future accurate measurements of $\BR (\Ktopinunu)$ and $\BR
(\KLtopinunu)$ close to their SM predictions is shown in the left
panel of \Fig{fig:fig2}. As can be seen the expected precision of this
determination of $(\bar{\rho}, \bar{\eta})$ can easily compete with
the one from the present global CKM fit \cite{Charles:2004jd}. A
comparison of $\sin 2 \beta$ determined from clean $B$-physics
observables with $\sin 2 \beta$ inferred from the $\Ktopinunus$ system
offers a very precise and highly non-trivial test of the CKM
picture. Both determinations suffer from very small theoretical errors
and any discrepancy between them would signal non-CKM physics, as
illustrated by the hypothetical example in the right panel of
\Fig{fig:fig2}. In particular, for $\BR  (\Ktopinunu)$ and $\BR 
(\KLtopinunu)$ close to their SM values, the reduction of the
theoretical error in $\Pc$ from $10.1 \%$ down to $2.4 \%$ translates
into the following uncertainties  
\cite{Buras:2005gr}     
\beq \label{eq:ckm}
\f{\sigma \left ( | V_{td} | \right)}{| V_{td} |} = 
\begin{cases}
\pm 4.1 \% \, , & \hspace{1mm} \text{NLO} \, , \\
\pm 1.0 \% \, , & \hspace{1mm} \text{NNLO} \, , 
\end{cases} \hspace{5mm}
\sigma \left ( \sin 2 \beta  \right ) = 
\begin{cases}
\pm 0.025 \, , & \hspace{1mm} \text{NLO} \, , \\
\pm 0.006 \, , & \hspace{1mm} \text{NNLO} \, , 
\end{cases} \hspace{5mm}
\sigma \left ( \gamma \right ) = 
\begin{cases}
\pm 4.9^\circ \, , & \hspace{1mm} \text{NLO} \, , \\
\pm 1.2^\circ \, , & \hspace{1mm} \text{NNLO} \, ,
\end{cases} 
\eeq
implying a very significant improvement of the NNLO over the NLO
results. Here $V_{td}$ is the element of the CKM matrix and $\beta$
and $\gamma$ are the angles of the UT. In obtaining these numbers we
have used $\sin 2 \beta = 0.724$ and $\gamma = 58.6^\circ$
\cite{Charles:2004jd} and included only the theoretical errors quoted
in \Eq{eq:Pc}. Obviously the determination of the CKM parameters from
the $\Ktopinunus$ system will depend on the progress in the
determination of $\mc (\mc)$ and the measurements of both branching
ratios. Also a further reduction of the error in $| V_{cb} |$ would be
very welcome in this respect.       

\section{\label{sec:conclusions}CONCLUSIONS}

An accurate measurement of $\BR (\Ktopinunu)$, either alone or
together with one of $\BR (\KLtopinunu)$, will provide a very
important extraction of the CKM parameters that compared with the
information from $B$-decays will offer powerful and crucial tests
of the CKM mechanism embedded in the SM and all its minimal flavor
violating extensions. The drastic reduction of the theoretical 
uncertainty in $\Pc$ achieved by the recent NNLO computation will play
an important role in these efforts and increases the power of the
$\Ktopinunus$ system in the search for new physics, in particular if
$\BR (\Ktopinunu)$ will not differ much from the SM prediction.   

\begin{acknowledgments}
We are grateful to A.~J.~Buras, M.~Gorbahn and U.~Nierste for fruitful
collaboration, and A.~H\"ocker for providing the plot in the left
panel of \Fig{fig:fig2}. This work was supported in part by the 
U.S. Department of Energy under contract No.~DE-AC02-76CH03000. 
\end{acknowledgments}


\end{document}